\documentstyle[epsf]{l-aa}
\begin{document}
\thesaurus{02.          
              (08.14.1;   
               02.04.1;   
               02.13.1;   
               02.18.6;   
               02.18.7)    
            }

\title{Model neutron star atmospheres with low magnetic fields}
\subtitle{I. Atmospheres in radiative equilibrium}

\author{V.~E. Zavlin\inst{1}, G.~G. Pavlov\inst{2,3}
and Yu.~A. Shibanov\inst{3}}

\institute{
Max--Planck--Institut f\"ur Extraterrestrische Physik,
 Giessenbachstrasse, D-85740 Garching, Germany \and
Pennsylvania State University, 525 Davey Lab, PA 16802, USA \and
Ioffe Institute of Physics and Technology, 194021,
St.~Petersburg, Russia}

\date{Received January. 1996; accepted March 30. 1996}

\maketitle

\newcommand{\gapr}{\raisebox{-.6ex}{\mbox{
$\stackrel{>}{\mbox{\scriptsize$\sim$}}\:$}}}
\newcommand{\lapr}{\raisebox{-.6ex}{\mbox{
$\stackrel{<}{\mbox{\scriptsize$\sim$}}\:$}}}
\def\d{{\rm d}}
\newcommand{\be}{\begin{equation}}
\newcommand{\ee}{\end{equation}}
\newcommand{\req}[1]{(\ref{#1})}
\newcommand{\dd}{{\rm d}}
\def\dy{{{\rm d}\over{{\rm d}{\it y}}}}
\def\dT{{{\rm d}\over{{\rm d}{\it T}}}}
\def\Tef{{\it T}_{\rm eff}}
\def\yef{y_{\rm eff}}

\begin{abstract}
We present a detailed investigation of neutron star atmospheres
with low magnetic fields, $B\lapr 10^8 -10^{10}$~G, which do not
affect opacities and equation of state of the atmospheric
matter. We compute the atmospheric structure, emergent spectral
fluxes and specific intensities for hydrogen, helium and
iron atmospheres in a wide domain of effective temperatures
and gravitational accelerations expected for neutron stars.
The iron atmospheres are computed with the opacities and
equations of state from the OPAL opacity library.
We show that the model atmosphere spectra are substantially
different from the blackbody spectra.  For light element
atmospheres, the flux is greater than the blackbody flux, 
and the spectrum is harder, at high photon energies,
whereas at low energies the spectral flux follows the Rayleigh-Jeans
law with a (surface) temperature lower than the effective temperature.
The spectra of iron atmospheres display prominent spectral
features in the soft X-ray range.
The emergent specific intensity is anisotropic,
with the anisotropy depending on energy. These properties
of the atmospheric radiation should be taken into account
for the proper interpretation of the thermal
component of the neutron star radiation detectable
at  X-ray through UV energies. In particular, fitting of
the thermal component with the blackbody model may result in
substantially incorrect parameters of the radiating regions. 
We discuss an application of our atmosphere models to the nearby
millisecond pulsar PSR J$0437-4715$.


\end{abstract}
\newpage

\section{Introduction}                                        
Thermal radiation from the surfaces of neutron stars (NSs) 
provides important information about their thermal histories
and the properties of the high-density matter in their
interiors (Pethick ~1992; Tsuruta 1995). The NS thermal-like
radiation has been observed recently in the soft X-ray range
with $ROSAT$ (\"Ogelman 1995; and references therein)
and $EUVE$ (e.~g., Foster et al.~1996; Halpern et al.~1996) 
and in the UV-optical range with $HST$ 
(Pavlov et al.~1996). For the proper
interpretation of these observations, one should take into
account that NSs are not perfect blackbody emitters ---
properties of their radiation are determined,
like  in usual stars, by the radiative transfer in their
atmospheres, and the emergent spectra can differ substantially
from the Planck spectrum, depending on the chemical composition,
magnetic field, energy flux and gravity in surface layers
(Pavlov et al. 1995, and references therein).

There are at least  two factors which make NS atmospheres
different from those of normal stars: the immense gravity
and huge magnetic fields. The first factor is important for
all NSs; it leads to strong compression and non-ideality
of the atmospheric matter, which affects considerably 
equation of state (EOS) and opacity. Even much greater
and more complicated effects on the opacity and EOS can be
expected from the very high magnetic field which, in particular, makes the NS
atmospheres essentially anisotropic.

First NS atmosphere models have been developed by Romani (1987; hereafter R87).
He employed (nonmagnetic) opacities from the Los Alamos Opacity 
Library (LAOL) to calculate atmosphere models and
emergent spectra  for a few effective
temperatures, $5.5 \leq \log\Tef \leq 6.5$, and chemical compositions
(pure helium, carbon, oxygen, iron, and a mixture of elements
with cosmic abundance)
and showed that the spectrum could indeed be quite different from the blackbody,
especially for the helium models. Further work has been concentrated
on NS atmospheres with ``typical'' (strong) NS magnetic fields, 
$B=10^{11}-10^{13}$~G,
which drastically change opacities and EOS (Miller 1992; Shibanov et al. 1992;
Shibanov et al. 1993; Ventura et al. 1993; Pavlov et 
al. 1994; Pavlov et al. 1995;  
Shibanov \& Zavlin 1995; Shibanov et al. 1995a; Zavlin et al. 1995a). 
These papers (except that of Miller) were devoted to
pure hydrogen NS atmospheres because no reliable opacities and EOS had
been developed for non-ideal partly ionized gases of heavier elements.
It has been demonstrated that the strong magnetic fields not only 
dramatically affect the emergent spectra, but they also lead to
remarkable anisotropy and strong polarization of the NS thermal radiation. 

Importance of the
magnetic field depends on how large is the electron cyclotron
energy, $h\nu_B=heB/m_ec$, in comparison with other characteristic
energies.  For instance, if we consider a completely ionized (hot) atmosphere,
where the main source of opacity is the free-free transitions
and/or scattering on electrons, the opacity is not affected by
the field for $h\nu_B\ll h\nu$, $k_BT$
(see, e.~g., Kaminker et al. 1983). Since the energy transfer in
the atmosphere is mainly driven by photons with $h\nu \sim (1-10) k_BT$,
one can neglect the magnetic effects if $B \ll (m_ec/he)k_BT \sim 10^{10}
(T/10^6\, {\rm K})$~G. If a substantial fraction of bound atoms or
ions is present, the magnetic effects are determined by the ratio
$h\nu_B/Z^2{\rm Ry} \sim (B/10^9~{\rm G}) Z^{-2}$, 
where $Z$ is a characteristic ion
charge, and ${\rm Ry}=13.6$~eV is the Rydberg energy. When this ratio
is $\lapr 1$, magnetic effects on the bound-bound transitions between lower
quantum levels, as well as on the bound-free transitions well above
photoionization thresholds, can be neglected. 
Although transitions which involve highly excited levels may be affected 
by much lower fields,
 they usually are not important for conditions typical for NS atmospheres. 

Thus, depending on atmospheric temperature and chemical composition,
one can neglect the magnetic effects on opacities and EOS if
$B\ll 10^8-10^{10}$~G. This means that there exists at least one class of 
NSs, millisecond pulsars, for which the magnetic effects can be expected
to be small.
Four millisecond pulsars have been detected in soft X-rays (Halpern 1996,
and references therein). 
The radiation from at least one of them --
a nearby, bright PSR J$0437-4715$, with the field $B\sim 3\times 10^8$~G --
has been shown to have a thermal-like component apparently originating from
the NS surface (Becker \& Tr\"umper 1993; Edelstein et al.~1995; Halpern
et al.~1996).  To interpret these data properly, the
observed spectra and the light curves should be compared to 
nonmagnetic atmosphere models. 
One cannot exclude also that high magnetic fields of
ordinary pulsars are concentrated in small ``platelets'', whereas a substantial
fraction of the NS surface has much lower fields 
(Page et al. 1995).
Finally, since the NS magnetic fields are subject to decay
(although characteristic time of this decay remains highly uncertain ---
see, e.~g., Bhattachrya 1995), one can expect
that very old NSs, most of which do not show pulsar activity, have
magnetic fields sufficiently low for  nonmagnetic atmosphere models
could be applied to analyze their thermal radiation. 
Detection of two old NS candidates has been reported recently
(Stocke et al.~1995; Walter et al.~1996).
Although identification of these objects as NSs requires further confirmation,
the amount  of dead pulsars observable through their thermal radiation
 should exceed  considerably that of active pulsars,
 so that one can expect more such objects to be discovered in near future. 

Thus, investigation of the nonmagnetic NS atmosphere models
can be justified not only by their relative simplicity ---
the models can be applied to analysis of observations of real
objects. Although a set of such models has been presented in R87,  
there exist several reasons to re-examine the previous results and 
to continue work on the nonmagnetic models.

(i)   New, much improved 
opacity codes have been developed
by the OPAL group, and a comprehensive set of the opacity data has
been computed (Iglesias et al. 1992; Rogers \& Iglesias 1994).
These new opacities substantially exceed those from 
LAOL for heavy elements just at the densities and temperatures
characteristic for the NS atmospheres.
Particularly interesting for the application to NS atmospheres
is a set of monochromatic opacities for pure iron composition,
generated at request of, and in collaboration with,
Roger Romani (Romani 1996, private 
communication). Rajagopal and Romani (1996; hereafter RR96;
see also Romani et al. 1996) have
used the results to develop modified NS atmosphere models.
\footnote{We became aware about RR96 after the present
paper had been generally completed.}  

(ii) No hydrogen atmosphere models and only three helium
models (for $\Tef=10^{5.5}$, 10$^6$ and 10$^{6.5}$) were presented 
in R87, and a few hydrogen and helium nonmagnetic models
were presented in our previous papers 
(Shibanov et al.~1992; Zavlin et al.~1994; Pavlov et al.~1995),
mainly for comparison  
with the magnetic models, without a systematic analysis.
However, the low-field NSs, including the 
millisecond pulsars, are assumed to have been passed,
during their long life, through an accretion  stage when  
a substantial amount of light elements have been accumulated at the NS surface.
(It may even be that
the surface magnetic fields are low because of accretion-induced
field decay --- see, e.~g., Bhattacharya 1995).  Due to gravitational 
stratification (Alcock \& Illarionov 1980) and/or
nuclear spallation reactions which destroy heavy elements
(Bildstein et al. 1992, 1993), 
the upper atmospheric layers can be expected to consist
mainly of lightest elements unless they have  been dragged downward
by convection or burned in nuclear reactions. Only two hydrogen
atmosphere models (for $\Tef =10^{5.5}$ and $10^6$ K) were presented in RR96,
which is not enough for a detailed analysis of how the atmospheric structure
and emergent spectra depend on the model parameters.

(iii) Only the (local) flux spectra of the emergent radiation have
been studied in both R87 and RR96.
However, the temperature distribution over
the NS surface should be, as a rule, nonuniform due to either
the presence of a hot polar cap in active pulsars (which is
demonstrated by the light curve of PSR J$0437-4715$ --- Becker
\& Tr\"{u}mper 1993) or anisotropy of the heat conduction in the
NS crust (Shibanov \& Yakovlev 1996). This means that the flux from
a NS  is to be determined by integration
of the angle-dependent specific intensities over the observable
NS surface; this flux may differ substantially from the local
flux due to the frequency-dependent limb-darkening effect which has not been
investigated for the nonmagnetic models. In addition, 
gravitational bending of the photon trajectories is to be taken into
account to obtain correct light curves of the NS thermal radiation
(Shibanov et al. 1995b; Zavlin et al. 1995b,~c).


(iv) It would be desirable to verify 
the models presented in R87 and RR96 using an alternative,
more modern numerical approach which has been applied by our group to modeling
magnetic NS atmospheres. In particular,  
that approach takes into account  anisotropy of radiation
in both solving the radiative transfer and calculating the 
atmosphere temperature distribution,
whereas the Lucy-Uns\"{o}ld method employed 
in R87 and RR96 for calculating the temperature distribution
is correct only in the
case of isotropic radiation field.  In addition, Thomson scattering,
treated in R87 with an  oversimplified approach,
and excluded from consideration in RR96,
may be important
for light element atmospheres at $\Tef\gapr 10^6$~K and is to be included
consistently in the atmosphere modeling.

In this paper we present a set of nonmagnetic NS atmosphere models 
of different chemical 
compositions computed with an advanced atmosphere modeling code
combined with the OPAL opacities.
We study the atmospheric structure, and the spectra and 
angular distribution of emergent radiation 
in a wide range of effective  temperatures and  surface gravities.
We consider the atmospheres  in radiative equilibrium,
which means that the total energy flux is transferred only
by radiation, whereas other heat transport mechanisms (convection, electron
heat conduction) are of no importance. 
The models with convection, which plays a role at lower $\Tef$,
 will be analyzed elsewhere (see Zavlin et al. 1996
for preliminary results).  Our estimates show that, 
for the models considered, contribution of  electron heat
conduction to the heat flux never exceeds a few tenths 
of percent at unit optical depth. It may reach several percent
only at very bottoms of heavy element atmospheres with lowest
effective temperatures; 
the intensities of far Wien tails of the emergent radiation 
formed at these layers
are too low to be of any practical importance. 
It should be mentioned that our estimates of the conductive
opacity exceed those of RR96 (making electron heat conduction less
important) because RR96 used an approximate equation with the
Coulomb logarithm omitted. On the other hand, the equations we
use (Hochstim \& Massel 1969) also become inaccurate at high
densities and low temperatures, so that the problem needs further
consideration.

We describe our approach to  atmosphere modeling in Section 2,
and present the results in Section 3.
Some observational implications are discussed in Section 4.   

\section{General description of the model}
\subsection{Basic equations}

Since a characteristic thickness of the NS atmosphere, $\sim 0.01- 1$~cm,
is much smaller than the NS radius,
and characteristic densities are rather high,
we consider the plane-parallel atmospheres in local thermodynamic equilibrium.
Following a standard approach (e.~g., Mihalas 1978), we 
proceed from the radiative transfer equation in the form of a second-order  
boundary problem: 
\be
\mu^2 \dy {1\over{k_\nu}} \dy u_\nu = k_\nu (u_\nu - S_\nu)~, 
\label{rte1}
\ee
\be
\frac{\mu}{k_\nu} \dy u_\nu = u_\nu\vert_{y=0},~~~~~
u_\nu=B_\nu\vert_{y\to\infty}~,
\label{bc1}
\ee
where $0\leq\mu\leq 1$ is cosine of the angle
$\theta$ between the normal
to the atmosphere and the wave vector,
$u_\nu =u_\nu(y,\mu)= {1\over{2}}(I_\nu^+ + I_\nu^-)$, ~$I_\nu^+=
I_\nu(y,+\mu )$ and $I_\nu^-=I_\nu(y, -\mu )$ are
the specific spectral intensities
of the outward-  
and inward-directed streams of radiation,
$y$ is the column density ($\dd y =-\rho\, \dd z$), 
and $B_\nu=B_\nu(T(y))$ is the Planck function. The monochromatic
opacity $k_\nu=k_\nu(y)=\alpha_\nu(y)+\sigma_\nu(y)$ is the sum
of the true absorption opacity $\alpha_\nu$ and scattering opacity
$\sigma_\nu$.  For the source function $S_\nu =S_\nu(y)$ we
use 
\be
S_\nu = (\sigma_\nu J_\nu + \alpha_\nu B_\nu) k_\nu^{-1}~,
\label{sf}
\ee
where $J_\nu=J_\nu(y)=\int_0^1 u_\nu\, {\rm d}\mu$ is the mean
intensity.  This form of the source function
implies the isotropic scattering approximation which proved
to be very accurate even in the
case when scattering dominates over true absorption (e.~g., Mihalas 1978).
The boundary conditions (\ref{bc1}) assume that there is no
incident radiation at the outer boundary $y=0$, and the photon energy density
tends to its equilibrium value at large depths. 

In the present paper we consider only atmospheres in radiative
equilibrium, where convection and electron heat conduction are
negligible, and the energy is transferred solely by radiation.
In this case the  radiative flux 
is independent of $y$ and coincides with the total 
flux, $F\equiv \sigma_{SB}\Tef^4$,  
\be
\int_0^\infty F_\nu \dd\nu = 4\pi \int_0^\infty \dd\nu \int_0^1 \dd\mu
\frac{\mu^2}{k_\nu} \frac{\dd u_\nu}{\dd y} = \sigma_{SB} \Tef^4~,
\label{te}
\ee
where $F_\nu(y)$ is the spectral flux, $\sigma_{SB}$ is the Stefan-Boltzmann
constant, and $\Tef$ is the effective temperature.
It follows from Eqs.~(\ref{rte1}) and (\ref{te}) 
that the radiative equilibrium condition can be also presented in the form
\be
\int_0^\infty  \alpha_\nu (J_\nu-B_\nu) \d\nu=0~.
\label{re}
\ee

We consider the atmospheres in hydrostatic equilibrium, which implies that
the gas pressure is $P = g\, y$, where $g$ is the gravitational acceleration  
(the radiative force is 
insignificant for the temperatures of interest, $\Tef \ll 10^7$~K). 
The above equations are to be supplemented with EOS,
$P=P(\rho, T)$, and true absorption and scattering opacities, 
$\alpha_\nu(\rho, T)$ and $\sigma_\nu(\rho, T)$. 
The opacities can be either extracted
from opacity libraries or calculated with the use of atomic data
(radiative cross sections). In the latter case, the set of 
equations of ionization 
equilibrium (EOIE) should be solved to find the fractions of ions
in different stages of ionization as well as the electron number density
$N_e(\rho, T)$. 

The input parameters of a model are the effective temperature $\Tef$,
surface gravitational acceleration $g$ (or the NS mass and
radius) and chemical composition. Solving these equations 
yields the atmospheric structure ($T(y)$, $\rho(y)$, $N_e(y)$, etc.),
and the radiative field in the atmosphere 
($F_\nu(y)$, $J_\nu(y)$, $u_\nu(y,\mu$), 
including the
angular and spectral distribution of the emergent radiation, 
$I_\nu(0,\mu) = 2 u_\nu(0,\mu)$).   

It should be noted that the models give the radiation and the 
atmospheric structure for a local
element of the NS surface. If the surface conditions are not
uniform (e.~g, the effective temperature may be higher at the magnetic poles),
the observable photon flux should be obtained from integration of
$I_\nu(0,\mu)$ over
the visible NS surface, with allowance for the gravitational redshift
and bending of the photon trajectories (e.~g., Shibanov et al. 1995b; 
Zavlin et al. 1995b,~c).

\subsection{Method of solution}

Solving the system of the model equations splits into two parts,
each of which can be solved with the knowledge of the other: calculation
of the radiation field given the structure of the atmosphere, and
calculation the structure given the radiation field.
We start from calculating  the temperature distribution in the Eddington 
approximation, integrating the equation 
\be
\dy \frac{T}{\Tef}=\frac{3}{16}k_R \left(\frac{\Tef}{T}\right)^{3}~
\label{edt}
\ee
with the boundary condition $T(0)=0.841\Tef$. 
Here $k_R=k_R(T,\rho)$ is the Rosseland
mean opacity which is calculated at each step of the integration, with $\rho(y)$
being found from EOS and hydrostatic equilibrium. The initial atmospheric
structure is used to obtain a first approximation for the radiation field.

Computing the radiation field starts from calculating the 
Eddington factors,
\be
f_\nu = J_\nu^{-1} \int_0^1\mu^2 u_\nu~\d\mu~~~~{\rm and}~~~~
h_\nu = J_\nu^{-1} \int_0^1\mu u_\nu~\d\mu~.  
\label{ef}
\ee
In deep layers, where the radiation is nearly isotropic 
and the diffusion approximation is valid, we have $f_\nu\simeq \frac{1}{3}$
and $h_\nu\simeq \frac{1}{2}$.
The dependences of $f_\nu$ and $h_\nu$ on $y$
account for the depth dependence of anisotropy of radiation. 
Integrating Eqs.~\req{rte1} and \req{bc1} over $\mu$ yields
the ordinary differential equation for the mean intensity, 
\be
\dy {1\over{k_\nu}} \dy f_\nu J_\nu = \alpha_\nu~(J_\nu - B_\nu)~,
\label{rte2}
\ee
with the boundary conditions
\be
\frac{1}{k_\nu} \dy f_\nu J_\nu = h_\nu J_\nu\vert_{y=0},
~~~~~J_\nu = B_\nu\vert_{y\to\infty}~.
\label{bc2}
\ee
Now $f_\nu(y)$ and $h_\nu(y)$ can be calculated by solving Eqs.~\req{rte1}, 
\req{rte2} and \req{ef} iteratively:
we start from solving Eq.~\req{rte1} with the source function 
$S_\nu=B_\nu$ to obtain
an initial approximation for $f_\nu$ and $h_\nu$ from Eq.~\req{ef};
then we use these $f_\nu$ and $h_\nu$ to solve \req{rte2},
substitute the solution  $J_\nu$ into the source function, and solve
Eq.~\req{rte1} to find new $u_\nu$ and, hence, $f_\nu$ and $h_\nu$, and
so on, until convergence of $f_\nu$ and $h_\nu$. Note that at each step
of these ``inner'' iterations we solve ordinary differential equations
instead of original integro-differential (with respect to $\mu$ and $y$)
Eq.~\req{rte1}.

In principle, one could use $J_\nu(y)$ obtained in the course of
the inner iterations to calculate a new atmospheric structure.
However, convergence of the ``global'' iterations can be significantly
improved (Auer \& Mihalas 1968) if instead we use 
the solution of the integro-differential (with respect to $\nu$ and $y$)
equation 
\begin{eqnarray}
\dy {1\over{k_\nu}} \dy f_\nu J_\nu = \alpha_\nu~(J_\nu - B_\nu) -
\label{rte3} \\
- \alpha_\nu \frac{\dd B_\nu}{\dd T}\, {{\int_0^\infty \alpha_{\nu'} 
(J_{\nu'} - B_{\nu'})~\dd \nu'}
\over{\int_0^\infty \alpha_{\nu'}(\dd B_{\nu'}/\dd T) \dd \nu'}}~,
\nonumber
\end{eqnarray}
with the boundary conditions \req{bc2} and with the Eddington factors
calculated as described above.
Since the spectral flux is
\be
F_\nu = \frac{4\pi}{k_\nu}\dy f_\nu J_\nu~,
\label{flux}
\ee
the additional (as compared to Eq.~\req{rte2}) integral term in Eq.~\req{rte3} 
automatically provides the total radiative flux $F$ to be constant throughout
the atmosphere at any given iteration.  Moreover, 
this integral term turns into zero identically when the radiation field
and atmospheric structure satisfy the radiative equilibrium condition \req{re}.

The mean intensity  found from Eq.~\req{rte3} is used to calculate
a new temperature  $T+\delta T$ for each atmospheric layer, with
the correction $\delta T$ equal to the ratio of the
two integrals in the additional term in Eq.~(\ref{rte3}).
Once the new $T(y)$ is obtained, we can complete calculation of the
atmospheric structure (and new opacities) making use of
the hydrostatic equilibrium,
EOS and EOIE, and then compute the corresponding radiation field as described
above (starting with calculation of the Eddington factors).
The global iterations continue until the maximum relative correction
$\vert\delta T/T\vert$ becomes smaller than a given value
(typically $\sim 1\%$).
Note that using the modified transfer equation (\ref{rte3}) substantially
accelerates the global iterations
(a reasonable accuracy is reached in 2--4 iterations).

We solve the model  equations  numerically by discretizing on
the depth variable $y$, photon energy $E = h \nu$ and
angular variable $\mu$. Typically, we use
a grid with  $100 - 150$ depth points logarithmically spaced
between $10^{-8} - 10^{-6}$ and $1 - 10$~g~cm$^{-2}$,
$100 - 200$ energy points
logarithmically spaced between $0.1\, k_B \Tef$ and $60\, k_B \Tef$, and
$50$ angular points for $\mu$ between 0 and 1.
A fourth-order Hermitian discretization method 
(Auer 1976) is applied to solve Eqs.~\req{rte2}, \req{bc2} and \req{rte3}.
We also use the elimination scheme suggested by 
Rybicki \& Hummer (1991), 
which provides a higher accuracy in comparison with the 
standard Gaussian method. 

Since the monochromatic opacities of heavy elements (particularly iron)
include many thousands of spectral lines, their frequency dependence
should be properly binned to use them for construction of the atmosphere
models. We average the original 
monochromatic opacities arithmetically
in each bin of the logarithmic frequency grid (typically $(1- 2)\times 10^3$
frequency points per bin) and use these binned opacities to compute
the atmospheric structure and a ``smoothed'' spectrum as described above.
We checked by varying the number of bins that $\sim 100-200$ bins
provide sufficient accuracy (a relative error $\lapr 1-2\%$) 
of the atmospheric structure and the total flux.
To obtain the spectrum with a fine spectral resolution, we calculate
the intensities of the outward- and inward-directed streams,
\be
I_\nu^+(y,\mu) = {1\over{\mu}}\int_y^\infty S_\nu
k_\nu\, \exp\left(
-{1\over{\mu}}\int_y^{y_1} k_\nu~\d y_2\right)~\d y_1
\label{intout}
\ee
and
\be
I_\nu^-(y,\mu) = {1\over{\mu}}\int_0^y S_\nu
k_\nu\, \exp\left(
-{1\over{\mu}}\int_{y_1}^y k_\nu~\d y_2\right)~\d y_1~,
\label{intin}
\ee
with the original monochromatic opacities (at $\sim 10^4$ frequency points).
The spectral flux is then computed as
\be
F_\nu(y) =2\pi\int_0^1 [I_\nu^+(y,\mu) - I_\nu^-(y,\mu)]\, \mu\, \dd\mu~.
\ee
These ``output'' spectra can be further smoothed, if needed, with
a standard spline procedure.

\subsection{Opacities and EOS}

The true absorption opacity  $\alpha_\nu$ is due to
the free-free, bound-free and bound-bound transitions;
$\sigma_\nu$ is dominated by the Thomson
scattering on electrons.
The main distinction of the opacities, EOS and EOIE
in a NS atmosphere from those in usual stars
is due to high non-ideality (pressure effects).
For modeling hydrogen/helium atmospheres, we calculate the opacities
and EOIE of non-ideal gases using the occupation probability
formalism (Hummer \& Mihalas 1988)
as described by Zavlin et al. (1994). The results
of these calculation agree fairly well with the OPAL
data (Iglesias et al. 1992) obtained with a quite
different approach (Rogers 1986).

For modeling iron atmospheres,
we use the opacities and EOS computed by the OPAL group,
in collaboration with R.~Romani.  The opacities $k_\nu$ were computed
for the domain $\log R = -5\, (1)\, 1$, $\log T = 4.5\, (0.25)\, 7.5$
($R \equiv\rho/T_6^3$, $\rho$ in g~cm$^{-3}$, $T_6 = T/(10^6~{\rm K})$).
They are provided on a grid of 
$10^4$ photon energies linearly spaced in the range
 ~$2\times 10^{-3}k_BT  \leq
 E \leq 20k_BT$. To obtain $k_\nu$ at given $T$ and 
$\rho$, we use the cubic spline interpolation of $\log k_\nu$
with respect to $\log R$ and $\log T$. 
For $ E >20  k_B T$ we use additional tables for the same $R$ and
$T$ and $1~{\rm eV} \leq  E \leq 10~{\rm keV}$.

\section{ Results}

We have computed a set of the atmosphere models for NSs with various
effective temperatures $\Tef$,
gravitational accelerations $g$ and chemical compositions.
In this paper we demonstrate representative models
for pure hydrogen, helium and iron compositions. 
The hydrogen and helium models are presented for 
$\log\Tef = 4.7\, (0.3)\, 6.5$; higher effective temperatures are not
expected to be observed in cooling NSs, and at lower $\Tef$
the helium atmospheres become convectively unstable. The
iron atmosphere models, where convection can develop at
higher $\Tef$, are presented for $\log\Tef = 5.3\, (0.3)\, 6.5$.
We chose a ``basic'' gravitational accelerations $g_{14}\equiv
g/(10^{14}$~cm~s$^{-2})=2.43$ (it  corresponds to standard NS mass  
$M = 1.4 M_\odot$ and radius $R=10$~km); a few examples are given
for $g_{14}=0.5$ and $4.5$.

\subsection{Atmospheric structure}

\begin{figure*}
\epsfxsize=16cm
\epsffile[10 70 800 450]{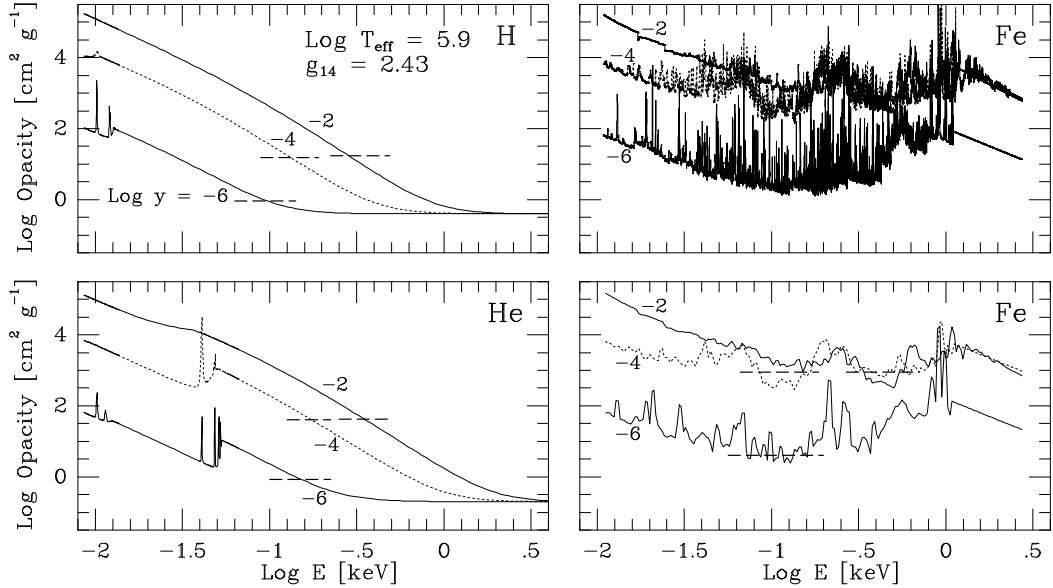}
\caption[ ]{
Monochromatic opacities {\em vs}.~photon energy at the layers
$y=10^{-6}$, $10^{-4}$ and $10^{-2}$ g cm$^{-2}$ of the hydrogen,
helium and iron atmospheres with $\Tef = 10^{5.9}$ K and
$g=2.43\times 10^{14}$ cm s$^{-2}$.
The upper right panel shows  the original Fe opacities ($10^4$
energy points); the binned opacities (200 points) are shown in
the bottom right panel. The horizontal dashed lines demonstrate
the Rosseland mean opacities $k_R$ at the corresponding layers.
}
\label{fig1}
\end{figure*}

The atmospheric structure has not been analized thoroughly in
the previous works; however, it is not only interesting in itself
--- it also allows one to better understand properties of
the emergent radiation.
Quite different structures of atmospheres composed of light
and heavy elements are due to  very different opacities.
An example of the energy dependence of the opacities at
the depths $\log y =-6$, $-4$ and $-2$ is shown in Fig.~1
for the hydrogen, helium and iron atmosphere
models with $\log\Tef =5.9$, $g_{14}=2.43$. We see that the
Fe opacities, dominated by numerous spectral lines,
are, on average, greater than H and He opacities, and their
mean slope at the most important frequency range is
almost flat.  On the contrary,  H and He opacities
are dominated by the free-free and bound-free transitions
($k_\nu\propto \nu^{-3}\propto E^{-3}$) at moderate and low frequencies
(energies), whereas at high frequencies they
are dominated by Thomson scattering (independent of $\nu$).
These distinctions are responsible for different atmospheric structures
and properties of the emergent radiation.

\begin{figure}
\epsfxsize=8.5cm
\epsffile[0 70 400 400]{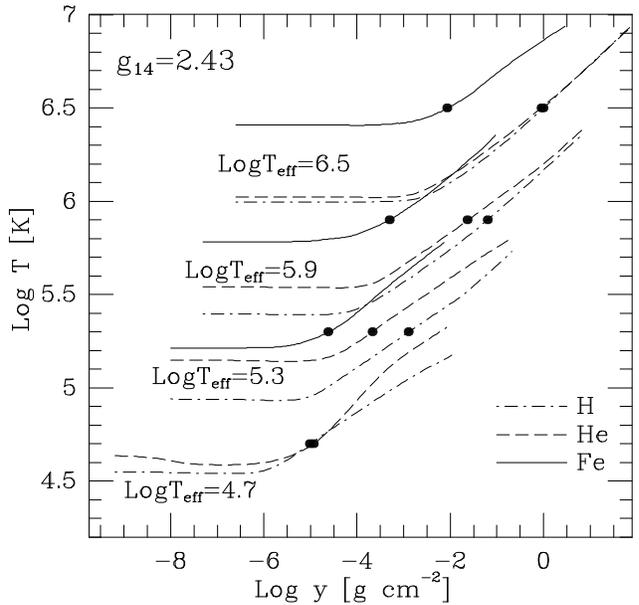}
\caption[ ]{
Temperature profiles $T(y)$ of the models with different effective
temperatures and chemical compositions. The filled circles mark the points
where $T(y)=\Tef$; the corresponding $y=\yef$ can be considered
as a characteristic depth of the atmosphere.
}
\label{fig2}
\end{figure}
%

\begin{figure}
\epsfxsize=8.5cm
\epsffile[0 70 400 400]{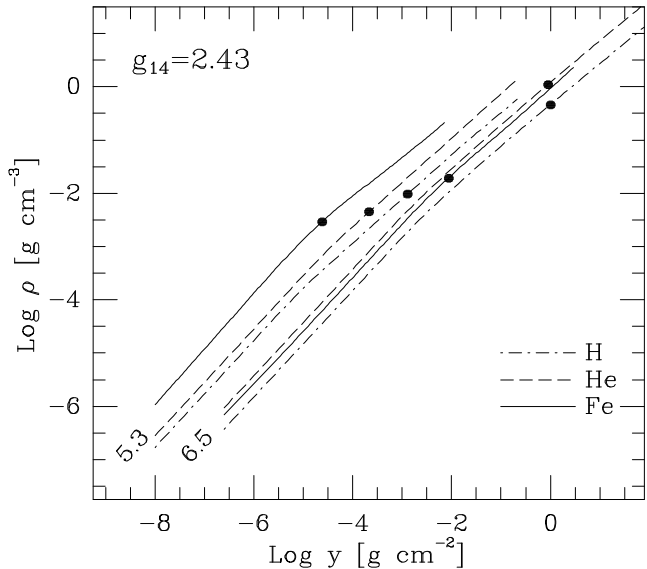}
\caption[ ]{
Mass density profiles $\rho(y)$ for two values of $\log \Tef = 5.3$
and $6.5$ and three chemical compositions.
  The filled circles indicate the characteristic
 densities $\rho_{\rm eff}= \rho(\yef )$ (cf. Fig.~2).
}
\label{fig3}
\end{figure}

\begin{figure}
\epsfxsize=8.5cm
\epsffile[0 70 400 600]{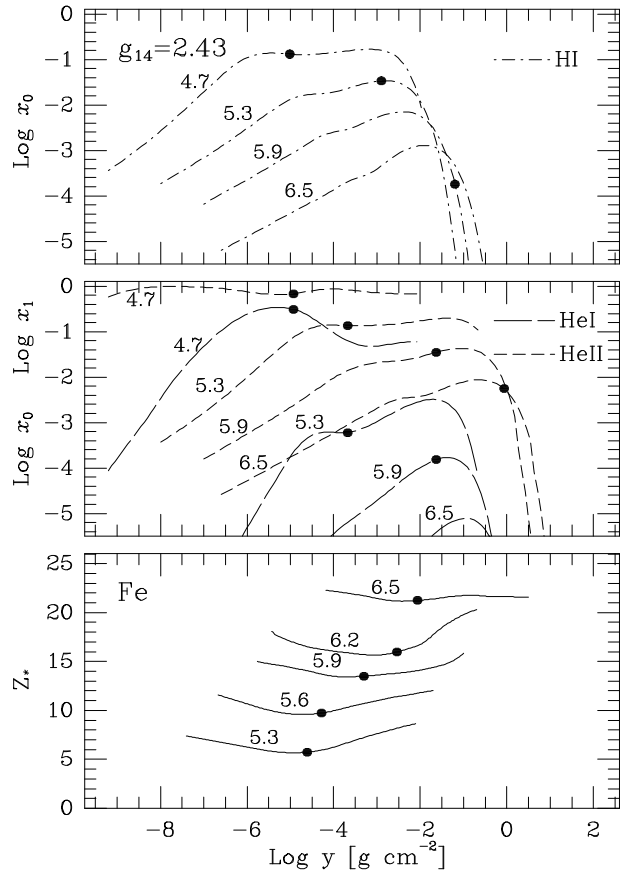}
\caption[ ]{
Depth dependences of ionization for
models with different values of $\log \Tef$ (numbers near the curves).
Two upper panels show profiles of non-ionized hydrogen
and helium, $x_0(y)$, and once-ionized helium, $x_1(y)$,
for pure hydrogen and helium models. The bottom panel
shows the depth dependences of the mean ion charge
$Z_\ast (y)$ for pure iron models. The filled circles
show the corresponding quantities at $y=\yef$.
}
\label{fig4}
\end{figure}

Particularly important component of the atmospheric
structure is the depth dependence of  temperature,
$T(y)$. Examples of $T(y)$ for several values of $\Tef$ and for 
three chemical compositions are demonstrated in Fig.~2.
We see that, for any $\Tef$, the temperature
at a given depth $y$ is systematically higher for iron atmospheres.
In other words, a characteristic  depth, $\yef$,
which can be defined by the equation $T(\yef)=\Tef$, is
smaller for iron atmospheres ($-4.6 <\log \yef < -2.1$
in our example) than for helium and hydrogen atmospheres
($-3.7 <\log \yef < 0$ and $-2.9 < \log \yef <0$,
respectively, for the same range of $\Tef$). Higher temperatures
for iron atmospheres at large depths, where
$T^4(y)\sim \Tef^4\tau_R = \Tef^4
 \int_0^y k_R\, \d y$, are due to higher Rosseland mean opacity $k_R$
(cf.~Fig.~1).
Substantially higher surface temperatures 
($T_s[\equiv T(0)]=(0.78 - 0.81)\Tef$
for Fe {\em vs}. $T_s=(0.31-0.53)\Tef$ for H, at $\log\Tef=5.3-6.5$)
can be explained by generally flatter frequency dependence
of the Fe opacity, which results in temperature profiles 
and surface temperatures close to those for the grey atmosphere
($T_s\simeq 0.811\Tef$).
The differences between the hydrogen and helium temperature
profiles depend on the degree of ionization in the atmosphere,
which, in turn, is determined by $\Tef$ and $g$. In particular,
at high $\Tef$ the light element plasma is completely ionized,
and its opacity in deep, dense layers is mainly determined by
the free-free transitions: $k_R\sim Z^3A^{-2}\rho T^{-7/2}$,
where $Z$ is the ion charge, and $A$ the atomic weight. 
Using this expression for $k_R$, Eq.~(\ref{edt}), and EOS for the ideal gas,
$\rho \propto A(Z+1)^{-1} g y T^{-1}$, 
we obtain $T^{15/2} \d T \propto g Z^{3} [A(Z+1)]^{-1}y\d y$,
which yields the following equation
 for the asymptotic behavior of the temperature of
a fully ionized atmosphere at large $y$:  
\be
T(y)\propto\left(\frac{Z^3}{Z+1}\frac{g}{A} \Tef^4\right)^{2/17} y^{4/17}~.
\label{tempas}
\ee
The value of the multiplier in front of $y^{4/17}$ for He exceed that
for H by 3\% only; this means that the temperature profiles
are essentially the same at sufficiently large depths, which we
observe for the models with $\log\Tef=6.5$. At $\log\Tef=5.9$,
the hydrogen atmosphere is completely ionized at the most important
depths, whereas the helium atmosphere contains a few percent 
of once-ionized helium, HeII (see Fig.~4), which substantially
increases the bound-free opacity (Fig.~1). This leads to
a greater difference between the He and H temperatures at
$y\sim \yef$. At $\log\Tef=5.3$ the difference is even more
pronounced because a larger fraction of HeII, $\gapr 10\%$, is available.
Finally, at $\log\Tef=4.7$, the H atmosphere contains $\sim 10\%$
of  neutral atoms HI, and the He atmosphere is mainly
composed of  HeII, with an admixture of HeI. In  both
cases opacity is dominated by the bound-free transitions,
and the temperature profiles are close to each other at $y\sim\yef$.

The temperature dependence of the He atmosphere with $\log \Tef =
4.7$ shows an interesting anomaly 
at small $y$: it slightly increases outward
at $\log y\lapr -7.5$. This inverse behavior 
is due to substantial growth with $y$ of the HeI fraction at
$\log y\lapr -7$ (see Fig.~4), accompanied by increasing role
of the  HeI photoionization absorption at $E\gapr 20- 25$ eV. 
To provide the radiative equilibrium (Eq.~(\ref{re})) at small $y$, 
the local blackbody intensity $B_\nu(T(y))$ is to be
greater than the mean intensity $J_\nu(y)$ at low energies (below the
main photoionization jump),
and smaller than $J_\nu(y)$ at high $E$ (above the jump).
At $\log y \lapr -8$ the main jump (provided by HeII) arises  at 
$\simeq 50-55$~eV,
whereas at $-7 \lapr \log y \lapr -6$ the jump is shifted to 
$\simeq 20-25$~eV
due to a greater fraction of HeI. Therefore,  for energies between 
approximately $20- 55$~eV, $B_\nu \gapr J_\nu$ at $\log y \lapr -8$,
while the opposite inequality holds at
$-7 \lapr \log y \lapr -6$. Since $J_\nu$ is practically
independent of $y$ in all these layers, the Planck function
at $\log y \lapr -8$ should be higher than at 
$-7 \lapr \log y \lapr -6$. This decrease
 of $B_\nu$ at these depths is possible 
if only the surface temperature $T(0)$ is higher 
than the temperature in the deeper layers. 

Figure 3 shows the depth dependence of the mass density $\rho(y)$
for the models with $\log\Tef=5.3$ and 6.5. The density
grows with $y$ linearly (exponentially with the geometrical depth $z$) 
at the upper (exponential) atmosphere, where the temperature 
is close to $T_s$. In deep layers, the density 
grows more gradually due to the steeper temperature growth. 
The density can be estimated as $\rho\sim
 m_H A (Z_\ast +1)^{-1} g y (k_B T)^{-1}$,
where $Z_\ast =Z_\ast (y) = N_e A m_H/\rho$ is the mean ion charge 
defined as the ratio of the electron number density $N_e$
to the summed number densities of atoms in all stages of
ionization. Since the hydrogen and
helium atmospheres at the temperatures considered are strongly 
ionized ($Z_\ast \simeq Z$), the density of a helium atmosphere
at a given $y$ exceeds that of a hydrogen atmosphere with the same
$\Tef$ and $g$.  In particular,
for deep layers of fully ionized atmospheres 
the curves $\rho(y)$ has a universal slope,
\be
\rho\propto Z^{-6/17} (Z+1)^{-15/17} A^{19/17} g^{15/17} 
\Tef^{-8/17} y^{13/17} 
\label{densas}
\ee
so that  $\rho^{(\rm He)}(y)/\rho^{(\rm H)}(y)\simeq 2.6$. 
Iron atmospheres at these temperatures are only partially ionized,
and the different disposition of the $\rho^{(\rm Fe)}(y)$ 
with respect to $\rho^{(\rm He)}(y)$ at $\log\Tef=5.3$ and 6.5
is explained by different degrees of ionization 
($Z_\ast\simeq 5-7$ and $21-22$,
respectively --- see Fig.~4).
Characteristic densities in different models, $\rho_{\rm eff}=
\rho(\yef )$, are generally lower for iron atmospheres,
$ \rho_{\rm eff}\sim 0.003 - 0.02$~g~cm$^{-3}$, than for hydrogen/helium
atmospheres, $\rho_{\rm eff}\sim 0.005 - 1$~g~cm$^{-3}$, for
the range $\log \Tef = 5.3 -6.5$. The ratio $z_{\rm eff} =
\yef /\rho_{\rm eff}$ gives a characteristic scale height
in an atmosphere; it varies from $\sim 0.01$~cm for Fe at $\log \Tef=
5.3$ to $\sim 2$~cm for H at $\log\Tef=6.5$.

The atmospheric opacity and, hence, the temperature, density and
properties of the emergent radiation depend essentially on the
ionization state at different depths. Figure 4 demonstrates the 
depth profiles of the fractions of ions in different stages
of ionization, $x_j=N_j/(N_0+N_1+\ldots +N_Z)$ ($j=0, 1,\ldots, Z$
is the ion charge). We show the nonionized fraction $x_0$ for
the hydrogen models, nonionized and once-ionized fractions,
$x_0$ and $x_1$, for the helium models,
and the mean ion charge $Z_\ast=\sum_{j=1}^Z j x_j$ for the iron models.  
The hydrogen atmospheres are strongly ionized at  the
considered temperatures (even at $\log\Tef = 4.7$
the nonionized fraction does not exceed 17\%);
however, the contribution of the bound-free transitions into
opacity is not negligible  at $x_0$ as small as $10^{-3}
-10^{-2}$ (cf.~Fig.~1).
At $\log y \lapr -2.5$ the values of $x_0$ are determined by
the usual Saha equation: $x_0$ grows with $y$ at very outer
layers, where the temperature is almost constant, due to
increasing density; the growth decelerates in deeper layers
due to increasing temperature. At large depths, 
where $\rho \gapr 10^{-2} -10^{-1}$~g~cm$^{-3}$,
the fraction $x_0$ decreases steeply due to pressure ionization.
Similar behavior is seen for the helium fractions $x_0$ and $x_1$,
although the ionization degree is lower than for hydrogen due to
higher ionization potentials. At moderate temperatures,
the main contribution to the
opacity comes from the bound-free transitions of HeII;
the non-ionized HeI plays a role at lower $\Tef$.
The iron atmospheres are only partly ionized even at high
$\Tef$ and very deep layers (e.~.g., $Z_\ast$ does not exceed
$ 21-22$ through the whole atmosphere with 
$\log\Tef = 6.5$) 
due to very high ionization potentials and binding energies.

The atmospheric structure depends also on the gravitational
acceleration $g$.
Our computations show that varying $g$ in the range expected for
the surfaces of NSs affects the temperature
profile only slightly: $T\propto g^{2/17}$ in very deep layers
(see Eq.~\req{tempas}), and the effect
on the surface temperature is even smaller. The characteristic depth
and density depend on $g$ much stronger; in particular, $\rho\propto
g$ in surface layers, and $\rho\propto g^{15/17}$ in deep layers.
The ionization degree is affected accordingly: generally, 
the ionized fractions grow with $g$, particularly
at the surface layers. 

\begin{figure*}
\epsfxsize=16cm
\epsffile[10 70 800 450]{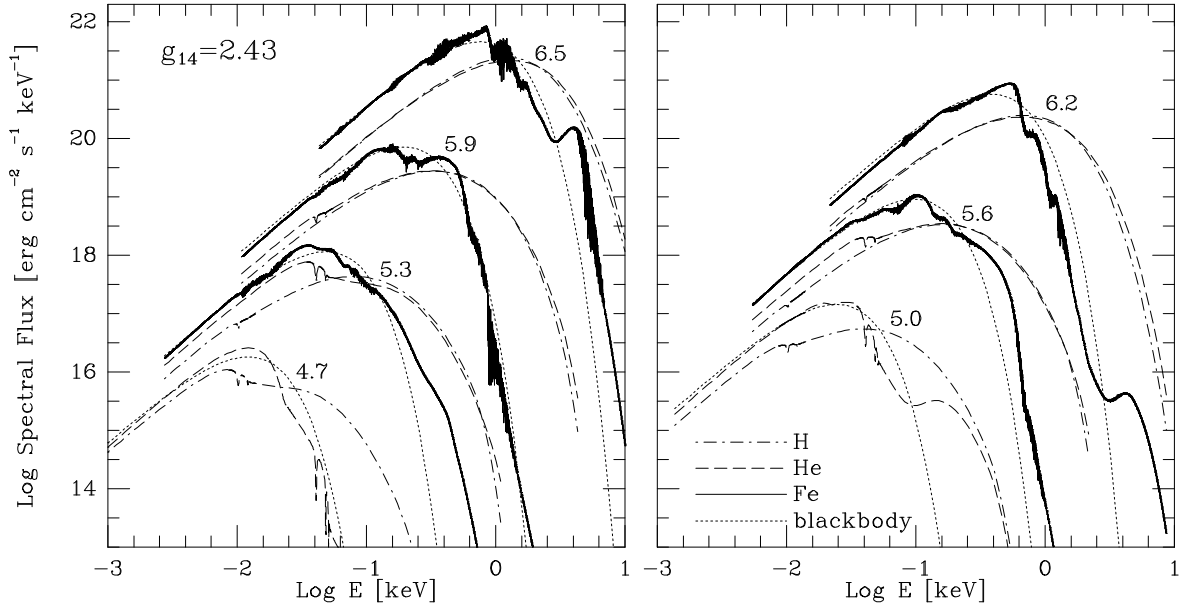}
\caption[ ]{
Spectral fluxes of emergent radiation for hydrogen, helium and iron atmospheres
 with different values of $\log \Tef$ (numbers near the curves).
Dotted curves show the corresponding blackbody fluxes $\pi B_\nu(\Tef)$.
}
\label{fig5}
\end{figure*}

\subsection{Spectral flux}

In Fig.~5 we show the spectral fluxes of the emergent radiation  
for  selected effective temperatures and 
three chemical compositions. The blackbody fluxes $\pi B_\nu(\Tef)$
are also shown for comparison.

We see that the model spectra are  substantially different
from the blackbody spectra with the same $\Tef$. The most striking
distinction is  much harder high-energy tails of the spectra
emergent from H and He atmospheres (see also   
 R87, Pavlov et al. 1995, and RR96). 
The reason for such behavior is 
that the H and He free-free and bound-free opacities rapidly 
decrease with increasing
photon energy ($\propto E^{-3}$ in a wide energy range --- see Fig.~1),
so that the radiation of higher energies
is formed in deeper and hotter layers. 
The maxima of the spectra of strongly ionized 
atmospheres (at $\log\Tef \gapr
5.0$ and 5.5 for H and He, respectively) lie at $E=(4.6-5.0) k_BT$,
i.~e., they are shifted by a factor of $1.6-1.8$
from the corresponding blackbody 
maxima towards higher energies.
Thomson scattering  dominates the opacities
at high energies (Fig.~1), which must be taken into account
for calculating the high-energy tails (i.~e., at $E \gapr 1$~keV for
the models with $\log \Tef = 5.9$).

At lower $\Tef$, the light element
atmospheres are less ionized, and  photoionization edges
and spectral lines are seen in their spectra.
In particular, the Lyman jump and Lyman-$\alpha$ line
of HeII become perceptible in the helium atmosphere spectra 
at $\log\Tef \lapr 6$, when the fraction of the once-ionized
helium is as small as a few percent (see Fig.~4).  The same
fraction of neutral hydrogen is needed to form the hydrogen Lyman
features which become well pronounced at $\log\Tef \lapr 5$.

\begin{figure}
\epsfxsize=8.5cm
\epsffile[0 70 400 550]{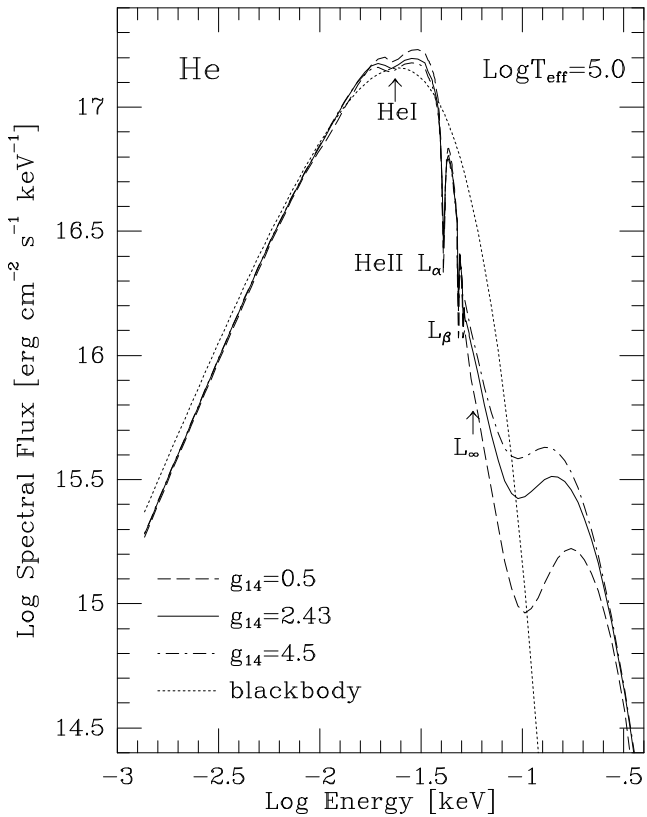}
\caption[ ]{
Spectral fluxes emergent from helium atmospheres with $\Tef =10^5$ K
and different surface gravities $g=g_{14}\times 10^{14}$ cm s$^{-2}$:
$g_{14}=0.5$ (dashed curve), $g_{14}=2.43$ (solid) and $g_{14}=4.5$
(dot-dashed).
}
\label{fig6}
\end{figure}

The low-energy tails ($E\ll k_BT$) of the model spectra 
follow the Rayleigh-Jeans law ($\propto  E^2$)
corresponding to the surface temperature $T_s (<\Tef )$,
i.~e., $F_\nu/\pi B_\nu = T_s/\Tef <1$. This occurs because
the opacities grow with decreasing $E$ so that the
low-energy radiation emerges from the very surface layers.
The low-energy flux is most strongly suppressed (by a factor
of up to $\simeq 3.2$) in hot H and He atmospheres because
of  smallness of the ratio $T_s/\Tef$ (see Fig.~2).

The spectra emitted from the iron atmospheres are much more
complicated, mainly due to numerous spectral lines and photoionization
edges produced by ions in various stages of ionization.
Generally, they are closer to the blackbody spectra,
and their high-energy tails are, on average,  softer  
than those emitted from light element atmospheres;
their low-energy tails lie only slightly below
the blackbody (Rayleigh-Jeans) spectra.
The reason is that the energy dependence of the iron opacities, 
in spite of its complexity, 
is, on average, flatter than that for H and He.
This makes the emergent spectrum to be
closer to the grey atmosphere 
spectrum which, in turn, only slightly differs from
the blackbody spectrum.
However, the local deviations from the blackbody spectra are quite
substantial, especially in the regions of the photoionization
edges, and strongly dependent on the effective temperature. 

Figure 6 demonstrates how the
gravitational acceleration affects the spectral flux emitted
by the helium atmosphere with $\log\Tef =5.0$. 
The effect is seen most vividly in the 
Lyman continuum of HeII ($E>54$~eV): the
Lyman discontinuity becomes progressively smoothed, and the
trough at $E\sim 0.1$~keV becomes shallower, with
increasing $g$. This occurs because the density grows
with $g$ 
 and the pressure effects  smooth
out the photoionization jump in the opacity spectra
(Zavlin et al. 1994).  The gravity effect (pressure broadening)
is also seen in the profiles of the absorption  lines, particularly  of
the 20~eV line of HeI. At higher temperatures, when
the light element atmospheres are strongly ionized
and the spectra are continuous, the gravity effect becomes insignificant.
In the case of iron composition
the effect of surface gravity can be seen at higher temperatures.

\begin{figure*}
\epsfxsize=16cm
\epsffile[10 70 800 450]{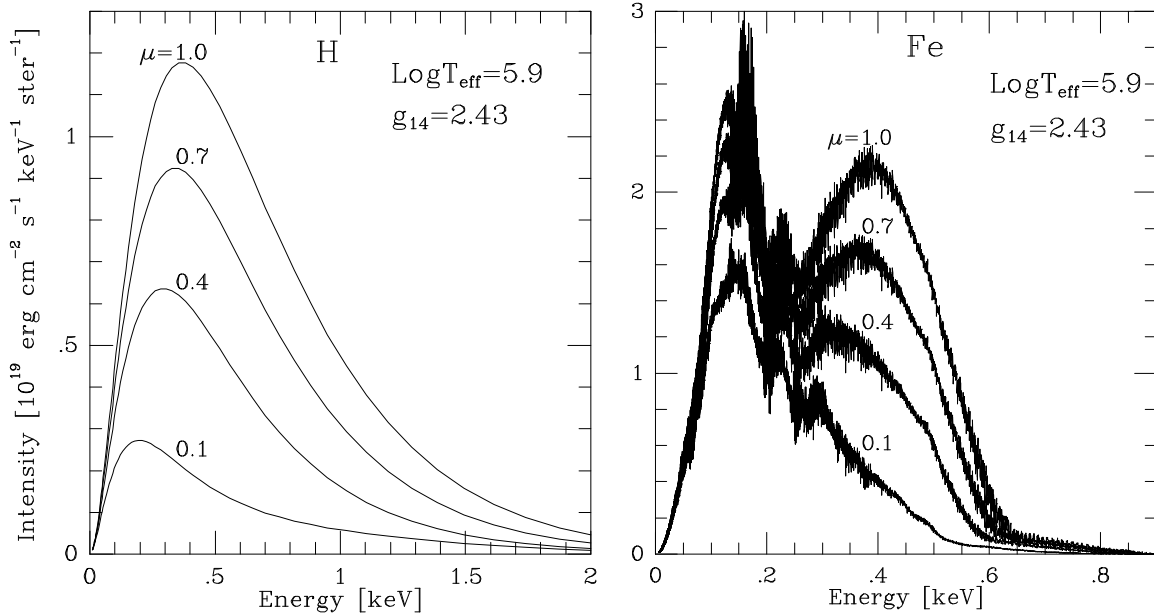}
\caption[ ]{
Spectra of the specific intensity of emergent radiation
for the hydrogen and iron atmospheres  with $\Tef = 10^{5.9}$ K
at different values of
$\mu = \cos \theta$ (numbers near the curves); $\theta$ is the angle
between the surface normal and direction of emission.
}
\label{fig7}
\end{figure*}

\subsection{Specific intensities: spectra and angular distributions}

Figure 7 demonstrates  spectra of the specific intensities 
of the emergent radiation at 
different angles $\theta=\arccos\mu$ between 
the direction of emission and the surface normal
for the models with hydrogen and iron compositions 
and the effective temperature of $10^{5.9}$~K.
We see that the intensity is rather anisotropic ---
it decreases with increasing $\theta$ (limb-darkening effect).
This is explained by the fact that radiation
emitted at larger $\theta$ escapes from shallower atmospheric
layers where the temperature is lower.
The anisotropy depends on photon energy
(the intensity spectrum varies with $\theta$)
and on chemical composition of the atmosphere. 
Note that the maximum of the
intensity spectrum shifts towards lower energies with increasing $\theta$.

\begin{figure}
\epsfxsize=8.5cm
\epsffile[0 70 400 400]{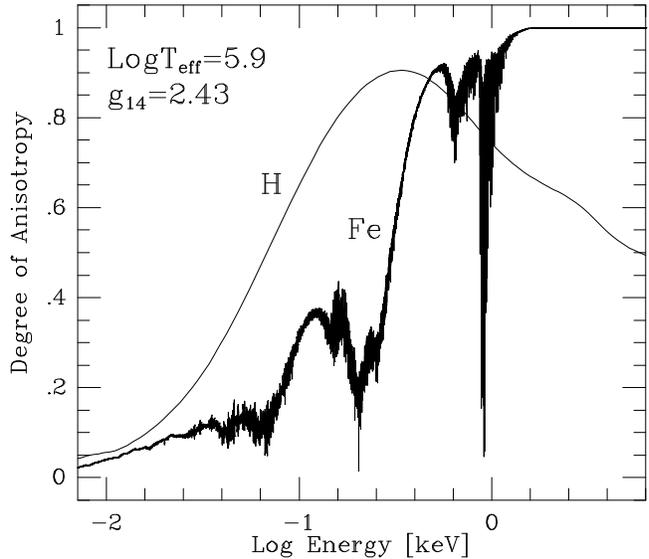}
\caption[ ]{
Degree of anisotropy $a_\nu$ (Eq.~\req{anisotr}) {\em vs}.~photon energy for
radiation emergent from hydrogen and iron atmospheres
with the same parameters as in Fig.~7.
}
\label{fig8}
\end{figure}

\begin{figure}
\epsfxsize=8.5cm
\epsffile[0 70 400 600]{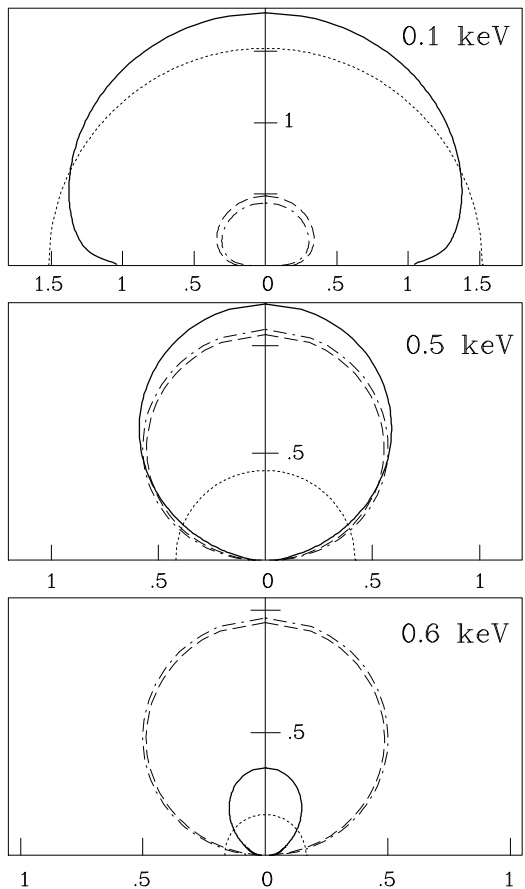}
\caption[ ]{
Polar diagrams of the specific spectral intensities
(in units of $10^{19}$ erg cm$^{-2}$ s$^{-1}$ keV$^{-1}$ ster$^{-1}$)
 emitted from hydrogen, helium and iron atmospheres (dash-dotted,
dashed and solid lines, respectively) with $\Tef =10^{5.9}$ K,
$g=2.43\times 10^{14}$ cm s$^{-2}$ at three values of photon
energy depicted at the upper-right corners of the panels.
The atmosphere normal is directed upward. The dotted lines
show the corresponding (isotropic) blackbody intensities.
}
\label{fig9}
\end{figure}

The energy-dependent degree of anisotropy  can be described
by the ratio 
\be
a_\nu = \frac{I_\nu(\mu=1) -I_\nu(\mu=0)}{I_\nu(\mu=1)+I_\nu(\mu=0)}~
\label{anisotr}
\ee
plotted in  Fig.~8;
it varies from 0 to 1 for atmospheres with the temperature
growing inward monotonously.  For the light element atmospheres, 
the anisotropy is small at low energies because the radiation
emerges from surface layers at any $\theta$. It grows with $E$
because radiation escapes from  hotter layers at $\theta=0$.
At very large $E$, when opacity is dominated by Thomson 
scattering, the angular dependence of the intensity is
determined by the Ambartsumian-Chandrasekhar function $H(\mu)$ (e.~g.,
Chandrasekhar 1960), and the degree of anisotropy is
$a_\nu=0.488$ (qualitatively, one can use 
the diffusion approximation, which yields $I_\nu(\mu) \propto  
1+1.5 \mu$ and $a_\nu \simeq 0.43$). 

In the case of iron atmospheres, the anisotropy is quite different.
In particular, it is very small in the regions of strong spectral
lines (e.~g., around 0.2 and 1~keV) 
which originate from the surface layers. (Similar reduction
of anisotropy near the spectral lines takes place for the light element
atmospheres at lower $\Tef$.) The radiation from the iron atmospheres
is very anisotropic ($a_\nu \to 1$) at high energies because
the opacity keeps decreasing with increasing $E$ (Thomson
scattering is negligible --- see Fig.~1).

Figure 9 shows representative angular distributions of the specific intensity
emergent from hydrogen, helium and iron atmospheres at different
photon energies. Although these  distributions do not show the prominent
features 
characteristic of radiation emitted from  strongly magnetized NS atmospheres
(see Fig.~4 in Shibanov \& Zavlin 1995),
this picture clearly illustrates strong dependences 
of the distributions on both the energy and chemical composition.

\section{Discussion}

The presented examples of the atmosphere models show that the
atmospheric structure and the properties of  emergent radiation
depend substantially on the chemical composition of radiating
layers and on the value of the effective temperature $\Tef$.
For the light element atmospheres
with a relatively high effective temperatures ($\log\Tef \gapr 5.3$
for hydrogen, $\log\Tef\gapr 6$ for helium), 
the atmospheric matter is strongly ionized, the 
(continuous) spectrum 
is much harder at X-ray energies than the 
blackbody spectrum, and the flux is lower than the
blackbody flux at low (UV/optical)
energies. For lower $\Tef$, spectral features arise in the
emergent spectra, particularly in the UV/EUV domain.
These distinctions mean that fitting thermal spectra observed 
from NSs with a blackbody model may be quite misleading.
In particular,
the blackbody fit of the {\em high-energy} (X-ray) 
tail of an observed spectrum would yield systematically
{\em overestimated effective temperature} and 
{\em underestimated} $A/d^2$ {\em ratio} ($A$ is the
area of the radiating surface, and $d$ is the distance).
The error introduced depends on the energy range
and response function of the specific detector; e.~g.,
for the $ROSAT$ PSPC the temperature may be overestimated
by a factor of up to $\simeq 3$, and $A/d^2$ underestimated by
more than an order of magnitude, at detectable NS temperatures.
On the contrary, {\em at low} (UV-optical) {\em energies} $\Tef$ is 
{\em underestimated}
or $A/d^2$ {\em overestimated}, typically by a factor of $2-3$.
To obtain reliable estimates of these quantities,
one should fit the results with various model atmosphere
spectra. It may happen that several models (e.~g., with
different chemical compositions) would  equally well fit the
data in a narrow spectral range. To avoid the ambiguity, multiwavelength
(e.~g., optical through X-ray) observations should be carried out, whenever
possible, and fitted with the atmosphere models.

Spectra of heavy element atmospheres are much more complicated
--- mainly because of numerous spectral lines,
as we see from the examples of iron atmospheres.
Note that although the shapes of our 
model spectra are qualitatively  similar to those obtained  in RR96,
the quantitative differences are significant, especially in 
domains around the strong photoionization features.
The difference may arise from both the different 
procedures for opacity binning
and different numerical approaches used for the modeling 
(e.~g., due to the inaccuracy of the Lucy-Uns\"{o}ld
temperature correction procedure used in RR96).
Although the overall shape of the iron atmosphere
spectra is not so drastically different from the blackbody,
using blackbody models for fitting in narrow energy intervals
may be even more misleading than for light element atmospheres 
because even the smoothed shape of the atmosphere spectrum depends
very sharply on $\Tef$. For instance, 
we see from Fig.~5 that 
increasing $\Tef$ by a factor of 2 completely changes
the shape of the iron atmosphere spectrum, and fitting the spectra
with the blackbody in a range of, e.~g., $0.2-0.6$~keV would
lead to quite different (and 
quite irrelevant) results.

We do not know a priori what is the chemical composition of
the NS surfaces. We can only expect that if there are light
elements, they should concentrate at the very surface,
but we cannot exclude that the light elements are virtually
absent, and the surface layers are composed of heavier elements.
To investigate the chemical composition, observations with higher spectral
resolution in the soft X-ray range would be very useful.
Our examples of the iron models show that 
the main spectral features (groups of lines and/or photoionization
edges) can be resolved with a quite moderate resolution, $\sim 3 - 30$,
which can be easily achieved with 
modern CCD detectors.
Thus, soft X-ray observations of NSs with future missions
($AXAF$, $XMM$, $ASTRO$-$E$) and their interpretation with the
NS atmosphere models will be able
to determine directly which elements reside at the NS surface.
 
Our results also show a very important property of the
atmospheric radiation --- it is anisotropic, with anisotropy
depending on frequency. This dependence  is different for 
different chemical compositions and effective
temperatures. This means that if
the temperature is not uniform over the NS surface (which is
very plausible for real NSs), one not only should take the anisotropy into
account to explain the light curves, but also one
cannot fit properly the spectra without allowance for anisotropy.
To calculate the flux as measured by a distant observer, one
should integrate the specific fluxes, $I_\nu \cos\theta$, 
over the visible NS surface with allowance for the gravitational
redshift and bending the photon trajectories. For instance, if
the radiation originates from a small hot spot (polar cap)
with a uniform $\Tef$ and chemical composition,
the flux as measured by a distant observer
is given by the equation (Zavlin et al. 1995c)
\be
F_\nu^{\rm obs} = \frac{A_a}{d^2}\, \frac{1}{(1+z)^{3}}\, I_{\nu_0}(\mu_c)~,
\label{obsflux}
\ee
where $A_a$ is the apparent area of the hot spot,
$z=[1-(2GM/c^2R)]^{-1/2}-1~$ is the gravitational redshift,
$\nu_0=(1+z)\nu$, $\mu_c$ is the cosine of the angle $\theta_c$
between the radius-vector of the spot center and wave vector of the photon
which travels to the observer along the bended trajectory.
  For  realistic NS radii, 
$R\gapr 5.6 (M/M_\odot)$~km ($z \lapr 0.45$), the apparent area
can be estimated as $A_a\simeq \pi R^2\gamma^2\mu_c$, where 
$\gamma$ is the angular size of the spot.
The important feature of Eq.~\req{obsflux} 
 is that the observed flux is proportional
not to the local flux, but to the specific intensity whose
spectrum (see Figs.~5 and 7) is substantially different
from that of the redshifted local flux used in RR96 (their Eq.~(7)).

The low-field atmosphere models can be directly applied to 
the nearby ($d\sim 140$~pc) millisecond pulsar PSR J$0437-4715$ 
($P=5.75$~ms) detected with the
$ROSAT$ PSPC (Becker \& Tr\"umper 1993) and $EUVE$ DS (Edelstein et al. 1995;
Halpern et al. 1996).
Fitting the $ROSAT$ spectrum of this low-field ($B\sim 3\times 10^8$~G)
pulsar with the single-component blackbody model proved to be
unsatisfactory --- the best-fit blackbody flux 
(corresponding to $T=1.5\times 10^6$~K) is lower
than the observed flux in
high-energy PSPC channels (above $\simeq 1$~keV).
However, the $ROSAT$ spectrum can be satisfactorily fitted
with double-component models. For the ``power-law + blackbody'' fit,
Becker \& Tr\"umper (1993) obtained
$T=1.7\times 10^6$~K and the radiating area $A=0.05$~km$^2$,
while Halpern et al. (1996)
obtained $T=(1.0-3.3)\times 10^6$~K and $A=0.008-1.1$~km$^2$.
In this model, the power law spectrum is of a nonthermal origin,
whereas the blackbody spectrum is produced by the hot polar
cap of the pulsar. It remains unclear with this interpretation 
why there is no a phase shift between the pulsations at low
energies (power law component) and high energies (thermal
component) --- being generated at different distances from
the NS surface, these components are expected to be phase-shifted
and to have different shapes. The values of the emitting area
are also uncomfortably low in comparison with that expected
from standard models for the radio pulsar polar cap,
$A=2\pi^2R^3/(cP) \sim 10$~km$^2$ (Arons 1981). 
Edelstein et al. (1995) ruled out that this two-component
fit is compatible with the $EUVE$ data, whereas Halpern et al. (1996)
came to the opposite conclusion and explained the $EUVE$ flux
with the power law component.

It is natural to expect
that fitting of the same data
with the model atmosphere spectra
and light curves would yield  quite different
results.
Indeed, our preliminary analysis of both the $ROSAT$ and $EUVE$ data
indicates that a single-component model, in which the soft X-ray radiation
originates from two polar caps of areas $A\simeq 2-3$ km$^2$
 covered with hydrogen or helium with
$\Tef\simeq (0.9 - 1.0)\times 10^6$ K,
fits satisfactorily
both the spectra and the light curves (Becker et al.~1996).
These results differ substantially from those of RR96 who fitted
the J0437-4715 spectrum with the local flux spectra, without
allowance for the limb darkening and gravitational lensing.
It worths noting that the  atmosphere models can be used to
predict the flux from the NS in the far-UV range ($\lambda
\lapr 2000-3000$~\AA), where it should overtake the flux of
the cold ($T\simeq 4000$~K) white dwarf companion. Measuring
this flux (e.~g., with $HST$) would be a critical test to
distinguish between different models of the radiating region.  Another
important test of the models
would be to observe the pulsar radiation
in the soft X-ray range  with a spectral resolution
sufficient to resolve spectral features; observations at moderate
X-ray energies ($\sim 1-10$~keV) would be useful to constrain the power law
component. Such observations would be quite feasible with
future X-ray missions ($SAX$, $AXAF$, $XMM$, $ASTRO$-$E$).
\acknowledgements
We thank F.~J. Rogers and C.~A. Iglesias for providing the
OPAL library data, J. Tr\"{u}mper for stimulating discussions. 
We are grateful to the referee, R.~W. Romani, whose remarks helped
us to improve this paper.
This work was partly supported by RFFI grant RBRF 96-02-16870A,
INTAS grant 94-3834 and NASA grant NAG5-2807. V.~E. Zavlin acknowledges
the Max-Planck fellowship.

\end{document}